\documentclass[aps,nofootinbib,twocolumn,floatfix,showpacs,preprintnumbers]{revtex4} 
\usepackage{graphicx}
\input epsf

\def\d{{\rm d}}

\def\lsim{\mathrel{\raise.3ex\hbox{$<$\kern-.75em\lower1ex\hbox{$\sim$}}}}
\def\gsim{\mathrel{\raise.3ex\hbox{$>$\kern-.75em\lower1ex\hbox{$\sim$}}}}

\def\cmm2{{\,\rm cm^{-2}}}
\def\cm2{{\,{\rm cm}^2}}
\def\cmm3{{\,{\rm cm}^{-3}}}
\def\gcmm3{{\,{\rm g\,cm^{-3}}}}

\def\fun#1#2{\lower3.6pt\vbox{\baselineskip0pt\lineskip.9pt
  \ialign{$\mathsurround=0pt#1\hfil##\hfil$\crcr#2\crcr\sim\crcr}}}

\def\be{\begin{equation}}
\def\ee{\end{equation}}
\def\bea{\begin{eqnarray}}
\def\eea{\end{eqnarray}}

\def\sigv{\langle\sigma v\rangle}

\begin{document}

\title{Extracting the Gamma Ray Signal from Dark Matter Annihilation \\in the Galactic Center Region}

\author{Scott Dodelson$^{1,2}$, Dan Hooper$^1$ and Pasquale D. Serpico$^1$}

\affiliation{$^1$Center for Particle Astrophysics, Fermi National
Accelerator Laboratory, Batavia, IL~~60510-0500}
\affiliation{$^2$Department of Astronomy \& Astrophysics, The
University of Chicago, Chicago, IL~~60637-1433}

\date{\today}
\begin{abstract}

The GLAST satellite mission will study the gamma ray sky with considerably greater exposure than its predecessor EGRET. 
In addition, it will be capable of measuring the arrival directions of gamma rays with much greater precision. These features each significantly enhance GLAST's potential for identifying gamma rays produced in the annihilations of dark matter particles. The combined use of spectral and angular information, however, is essential if the full sensitivity of GLAST to dark matter is to be exploited. In this paper, we discuss the separation of dark matter annihilation products from astrophysical backgrounds, focusing on the Galactic Center region, and perform a forecast for such an analysis. We consider both point-like and diffuse astrophysical backgrounds and model them using a point-spread-function for GLAST.  While the results of our study depend on the specific characteristics of the dark matter signal and astrophysical backgrounds, we find that in many scenarios it is possible to successfully identify dark matter annihilation radiation, even in the presence of significant astrophysical backgrounds.

\end{abstract}
\pacs{95.35.+d; 95.85.Pw}
\preprint{FERMILAB-PUB-07-632-A}
\maketitle

\section{Introduction}

Dark matter constitutes an essential element of modern cosmology. Evidence of its existence come from a wide variety of observations including the rotational speeds of galaxies~\cite{rotationcurves}, 
the orbital velocities of galaxies within clusters~\cite{clusters}, gravitational lensing~\cite{lensing}, 
the cosmic microwave background~\cite{wmap}, the light element abundances~\cite{bbn} and large scale 
structure~\cite{lss}. But despite these many observational indications of dark matter, we remain ignorant of the particle nature of this substance. To reveal the identity of dark matter, it will be crucial to measure its non-gravitational couplings. Efforts in this direction
include direct detection experiments, which are designed to observe the elastic scattering of dark matter 
particles with nuclei, and indirect detection experiments, which aim to detect the annihilation products 
of dark matter. Such annihilations could potentially produce observable fluxes of gamma rays, 
neutrinos~\cite{neutrinos}, positrons~\cite{positron}, antiprotons~\cite{antiproton}, 
antideuterons~\cite{antideu}, synchrotron radiation~\cite{syn} and X-rays~\cite{Bergstrom:2006ny}.
Of these various channels, gamma rays have the important advantage of retaining directional information. This is a feature that can be used, together with the peculiar energy spectra expected from dark matter annihilations, to disentangle dark matter annihilation products from astrophysical backgrounds~\cite{angular}.

In this paper, we study the ability of gamma ray experiments to identify dark matter annihilation radiation from the Galactic Center region (for earlier work on this subject, see Refs.~\cite{gchist}) by using both spectral and angular information. This is a goal of existing Atmospheric Cerenkov Telescopes (ACTs) including HESS~\cite{HESSurl} , 
MAGIC~\cite{MAGICurl}, VERITAS~\cite{VERITASurl} and CANGAROO-III~\cite{CANGAROOIIIurl},
 as well as of the forthcoming satellite mission, GLAST~\cite{GLASTurl}. Motivated by its sensitivity to gamma rays over an energy range well suited to dark matter searches (0.1-300 GeV), we focus here on the GLAST experiment.

The remainder of this paper is structured as follows: In Secs.~\ref{backgrounds} and~\ref{darkmatter} we describe the model used for the point-like 
and diffuse astrophysical backgrounds and the expected characteristics of a dark matter annihilation signal, respectively. In Sec.~\ref{results} we describe our analysis method and assess the ability of GLAST to either place limits on or identify dark matter annihilations in the Galactic Center region. In Sec.~\ref{measure} we extend our method to assess GLAST's ability to measure the annihilation cross section, mass and distribution of dark matter. In Sec.~\ref{conclusions} we summarize and present our conclusions. 
\section{Modeling the Galactic Center Backgrounds} \label{backgrounds}

The Galactic Center is a complex region of the sky at all wavelengths, the gamma-ray window being no exception. In this section, we discuss how, in our analysis, we treat the backgrounds for dark matter searches due to 
known and unknown astrophysical sources of gamma rays.

The first of the backgrounds we consider is the relatively bright, very high-energy gamma ray source observed 
by HESS~\cite{hess}, MAGIC~\cite{magic}, WHIPPLE~\cite{whipple} and CANGAROO-II~\cite{cangaroo}. This source is consistent with point-like emission and is located at $l = 359^\circ 56^\prime 41.1^{\prime\prime}\pm 6.4^{\prime\prime}$ (stat.), 
$b = -0^\circ2^{\prime}39.2^{\prime\prime}\pm 5.9^{\prime\prime}$ (stat.) with a systematic pointing error of
28$^{\prime\prime}$ \cite{van Eldik:2007yi}. It appears to be coincident with the position of Sgr A$^{\star}$, the black hole constituting 
the dynamical center of the Milky Way. The spectrum of this source is well described by a power-law
with a spectral index of $\alpha=2.25 \pm 0.04 (\rm{stat}) \pm 0.10 (\rm{syst})$ 
over the range of approximately 160 GeV to 20 TeV. Although speculations were initially made that this source could 
be the product of annihilations of very heavy ($\sim$10 to 50 TeV) dark matter particles~\cite{actdark}, this interpretation is disfavored by the power-law form of the observed spectrum. The source of these gamma rays is, instead, likely an astrophysical 
accelerator associated with our Galaxy's central supermassive black hole~\cite{hessastro}. In our analysis, we treat this source 
as a background for dark matter searches (see also Ref.~\cite{gabi}). 

Following the measurements of HESS, we describe the spectrum of this source as a power-law given by:
\begin{equation}
\Phi^{\rm ACT} = 1.0 \times 10^{-8} \left({ {E_{\gamma}}\over{{\rm GeV}}}\right)^{-2.25} {\rm GeV}^{-1} \, {\rm cm}^{-2} \, {\rm s}^{-1}.
\end{equation}
At energies below $\sim 200$ GeV, the spectrum of this source has not yet been measured. GLAST, however, will be capable of measuring the spectrum of this source at energies below the  thresholds of HESS and other ACTs. As any signal from dark matter annihilations is expected
 to be spatially extended, as a result of the dark matter halo profile (see Sec.~\ref{darkmatter}), GLAST's spectral measurement in the angular window around the source could be used to obtain a relatively pure determination of the background spectrum from the HESS source. This approach would be less effective, however, in the case in which the spatial distribution of dark matter annihilations is highly concentrated, such as might occur for a highly adiabatically contracted halo profile~\cite{ac} or for a density spike resulting from adiabatic accretion of dark matter onto our galaxy's supermassive black hole~\cite{spike}. In any case, the spectral index of this source has been measured by HESS at energies well beyond those expected to be relevant to data matter searches. Unless the spectral index changes significantly over the energy range of interest to GLAST, this information can be used to effectively predict the spectrum GLAST will detect from this astrophysical source. 

In addition to the HESS source, a yet unidentified source has been detected by EGRET approximately 0.2$^{\circ}$ away from the dynamical center of our galaxy~\cite{dingus,pohl}. Although the spectrum of this source is not yet well measured, its spectral index appears to be similar to that of the HESS source ($\alpha \approx 2.2$). We model the flux from this source as:
\begin{eqnarray}
\Phi^{\rm EG} &=& 2.2 \times 10^{-7} \,\left({ {E_{\gamma}}\over{{\rm GeV}}}\right)^{-2.2} \, {\rm GeV}^{-1} \, {\rm cm}^{-2} \, {\rm s}^{-1}  \nonumber \\
&\times& \exp(-E_{\gamma}/30 \, {\rm GeV}).
\end{eqnarray}

EGRET detected the presence of this source at energies up to $\sim 10$ GeV. If its power-law spectrum continues up to energies detectable by HESS and other ACTs, they would have observed the source as well. As they do not, we are forced to introduce a cutoff in the spectrum of this source. The precise value of the cutoff (30 GeV) was chosen arbitrarily. Once again, GLAST will measure the spectrum of this and other point sources and will thus not have to rely on such extrapolations or speculations in an actual analysis.

Even ideal point-like sources appear to be slightly extended in a realistic experiment due to the finite
angular resolution of the detector. To account for this, we model the GLAST point spread function by:
\begin{equation}
\frac{\d {\mathcal P}}{\d \Omega}(E_{\gamma},\theta)=\frac{1}{2\pi\,\delta^2(E_{\gamma})} \exp\bigg[-\frac{\theta^2}{2 \, \delta^2(E_{\gamma})}\bigg]\label{PSF},
\end{equation}
where $\theta$ is the difference between the measured and actual directions of the observed gamma ray and the solid angle element is $\d\Omega=\theta\d\theta\d\phi$, $\phi$ being the azimuthal direction of the displacement with respect to an arbitrary azimuth. The function $\delta(E_\gamma)$ is the angle within which 68\% of the gamma rays are reconstructed and is given in degrees by:
\begin{equation}
\log_{10} [\delta(E_{\gamma})] \approx -0.6-0.8 \log_{10} [E_{\gamma}/1\, {\rm GeV}].\label{PSF2}
\end{equation}
Strictly speaking, the normalization factor in Eq.~(\ref{PSF}) is correct only in the planar limit  
$\sin\theta\sim\theta$, that is when $\delta\ll\pi$, which for the case at hand is an excellent
approximation, see Eq.~(\ref{PSF2}). In this limit, the integral of Eq.~(\ref{PSF}) over 
the whole solid angle is equal to unity.

Finally, we consider the diffuse background of gamma rays in the region surrounding the Galactic Center. Unlike in the case of point sources, the detailed angular distribution of the diffuse background is not known and thus cannot be easily separated from a potential dark matter annihilation signal using angular information alone. In our analysis we consider two extreme possibilities in this regard. In the first case, we use a flat distribution for the diffuse background, with the same intensity in each angular bin (over a $2^{\circ}\times 2^{\circ}$ window centered around the Galactic Center). In the second case, we use an angular distribution which matches that of the dark matter signal. The latter case represents the extreme limit where there is no angular information with which to separate the annihilation signal from the diffuse background. Although the most realistic situation is bracketed by these two limits, we expect the uniform background case to be closer to reality than the latter one.  

The spectrum of the diffuse background is also not known in advance, and may be difficult to separate from dark matter annihilation radiation. Observations by HESS of the diffuse emission from the Galactic Center Ridge~\cite{ridge} 
find a spectrum in the region of the sky given by $-0.8^{\circ} < l < 0.8^{\circ}$, $-0.3^{\circ} <b < 0.3^{\circ}$ which can be described by a power-law:
\begin{equation}
\Phi^{\rm diff}_{\rm HESS} \approx 1.3 \times 10^{-4} \left({ {E_{\gamma}}\over{{\rm GeV}}}\right)^{-2.29} {\rm GeV}^{-1} \, {\rm cm}^{-2} \, {\rm s}^{-1} \, {\rm sr}^{-1}.
\end{equation}
The HESS observations do not, however, constrain the diffuse spectrum (or angular distribution) at energies below $\sim$200 GeV, which are of the most interest to GLAST. 
In our analysis, we assume the general parameterization 
\begin{equation}
\Phi^{\rm diff}(A,\alpha) = A \left({ {E_{\gamma}}\over{{\rm GeV}}}\right)^{-\alpha} {\rm GeV}^{-1} \, {\rm cm}^{-2} \, {\rm s}^{-1} \, {\rm sr}^{-1}\,,
\end{equation}
where $\alpha$ is allowed to vary between 1.5 and 3.0. We adopt an overall normalization, $A$, such that the integrated flux
of the diffuse background between 1 GeV and 300 GeV in a $2^\circ\times 2^\circ$ field of view around the Galactic Center
is equal to $10^{-4} \, {\rm cm}^{-2} \, {\rm s}^{-1} \, {\rm sr}^{-1}$. We do not, however, assume that this normalization is known in our analysis, leaving open the possibility that some of the diffuse gamma rays observed are the product of dark matter annihilations.

\begin{figure}[!htb]
\resizebox{7.0cm}{!}{\includegraphics{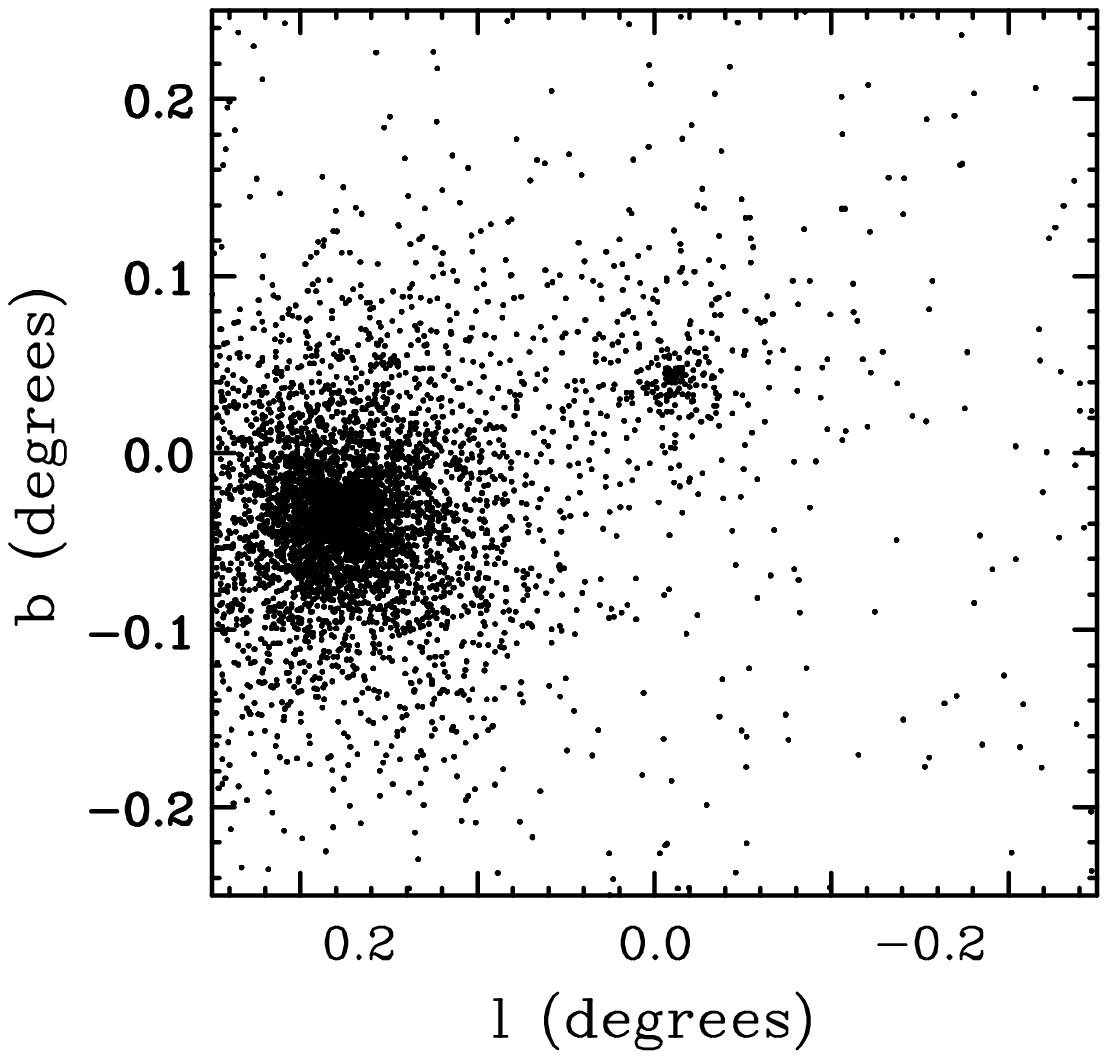}} \\
\vspace{0.5cm}
\resizebox{7.0cm}{!}{\includegraphics{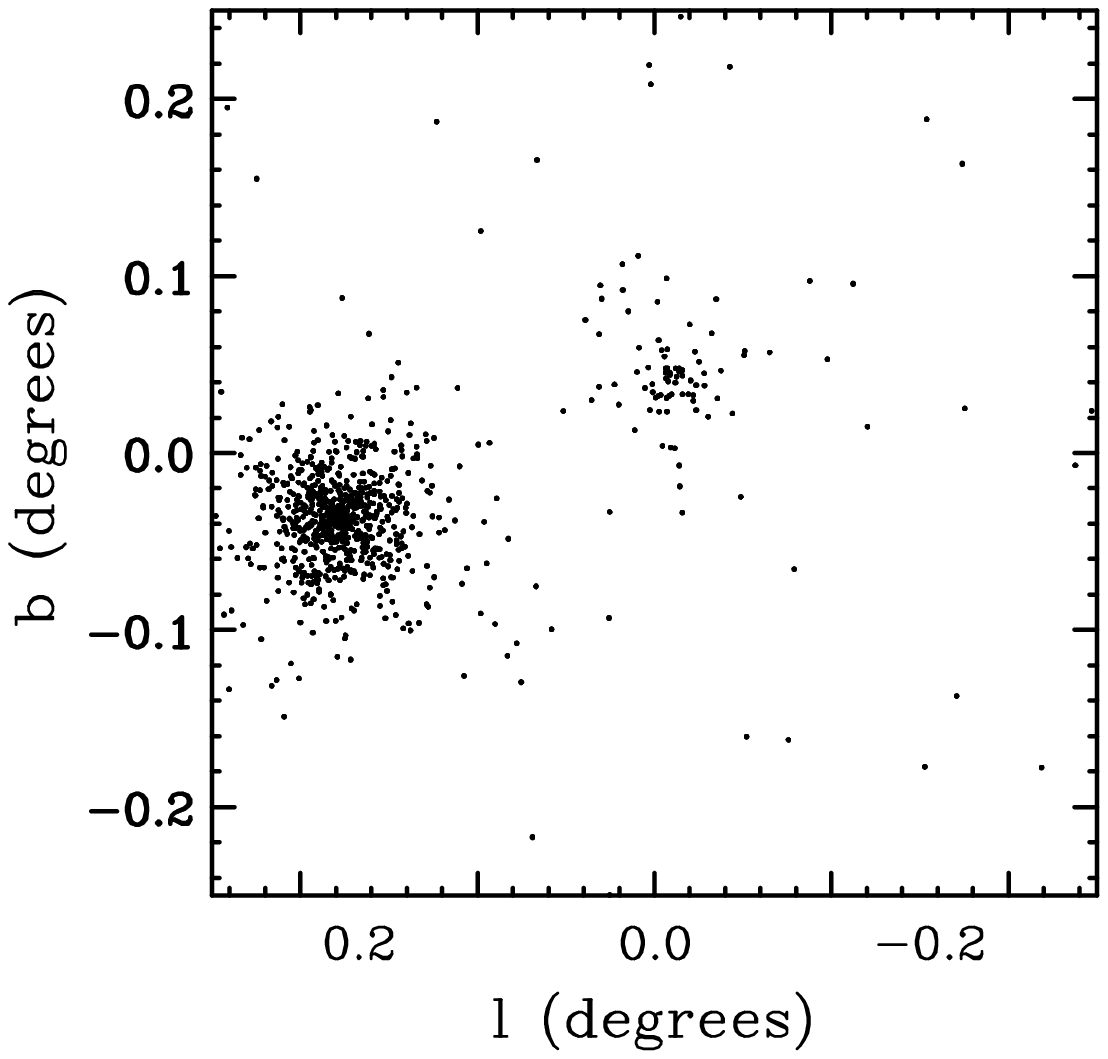}}
\caption{A simulated sky map of the gamma ray backgrounds present in the region of the Galactic Center, after two years of observation by GLAST. Each point denotes one gamma ray detected. In the top and bottom frames, photons with energy above 3 and 10 GeV are shown, respectively.
}
\label{bgmap}
\end{figure}

In Fig.~\ref{bgmap}, we show an example of the simulated backgrounds (including Poisson noise) as might be observed by GLAST in two years of observation. For GLAST, we have used a constant effective peak area above 1 GeV of $A_{\rm eff}=8.5\times 10^3\,$cm$^2$. We have also assumed that the Galactic Center will be within GLAST field-of-view  50\% of the time and that the reduction in effective
area for sources which are not located at the instrument zenith gives a mean
effective area equal to 60\% of the peak area~\cite{glast2}. We quantify the latter two effects by multiplying $A_{\rm eff}$ by the exposure parameter $\epsilon$, with $\epsilon=0.5\times 0.6=0.3$. The two distinctive features seen in the figure correspond to the HESS and EGRET sources described above. Also included is a diffuse background, distributed isotropically, with a spectrum given by $\Phi^{\rm diff} = 1.4 \times 10^{-4} (E_{\gamma}/{\rm GeV})^{-2.4} \, {\rm GeV}^{-1} \, {\rm cm}^{-2} \, {\rm s}^{-1} \, {\rm sr}^{-1}$.

\section{Gamma Rays From Dark Matter Annihilations}\label{darkmatter}

\begin{figure}
\resizebox{8.0cm}{!}{\includegraphics{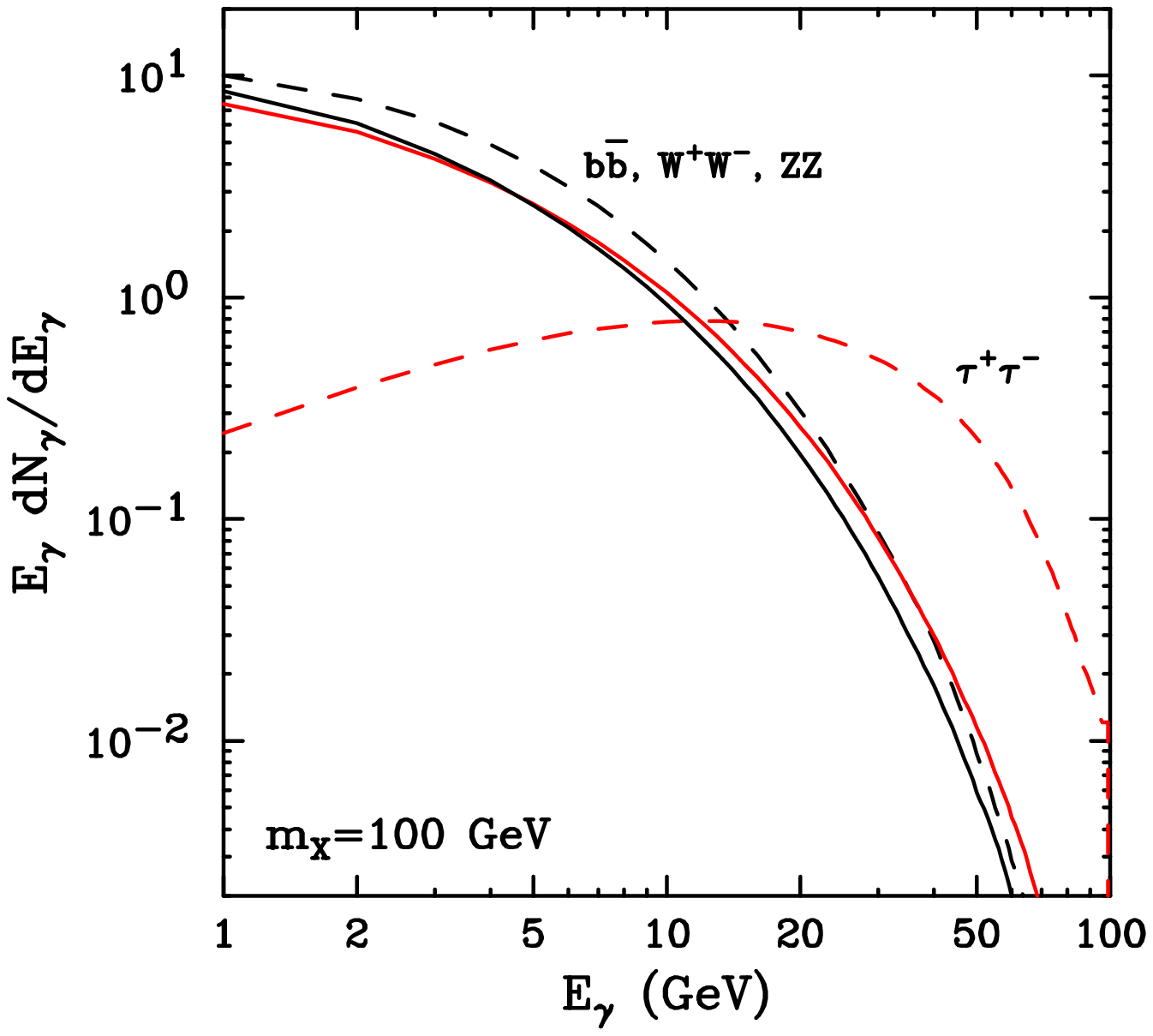}} \\
\vspace{0.6cm}
\resizebox{8.0cm}{!}{\includegraphics{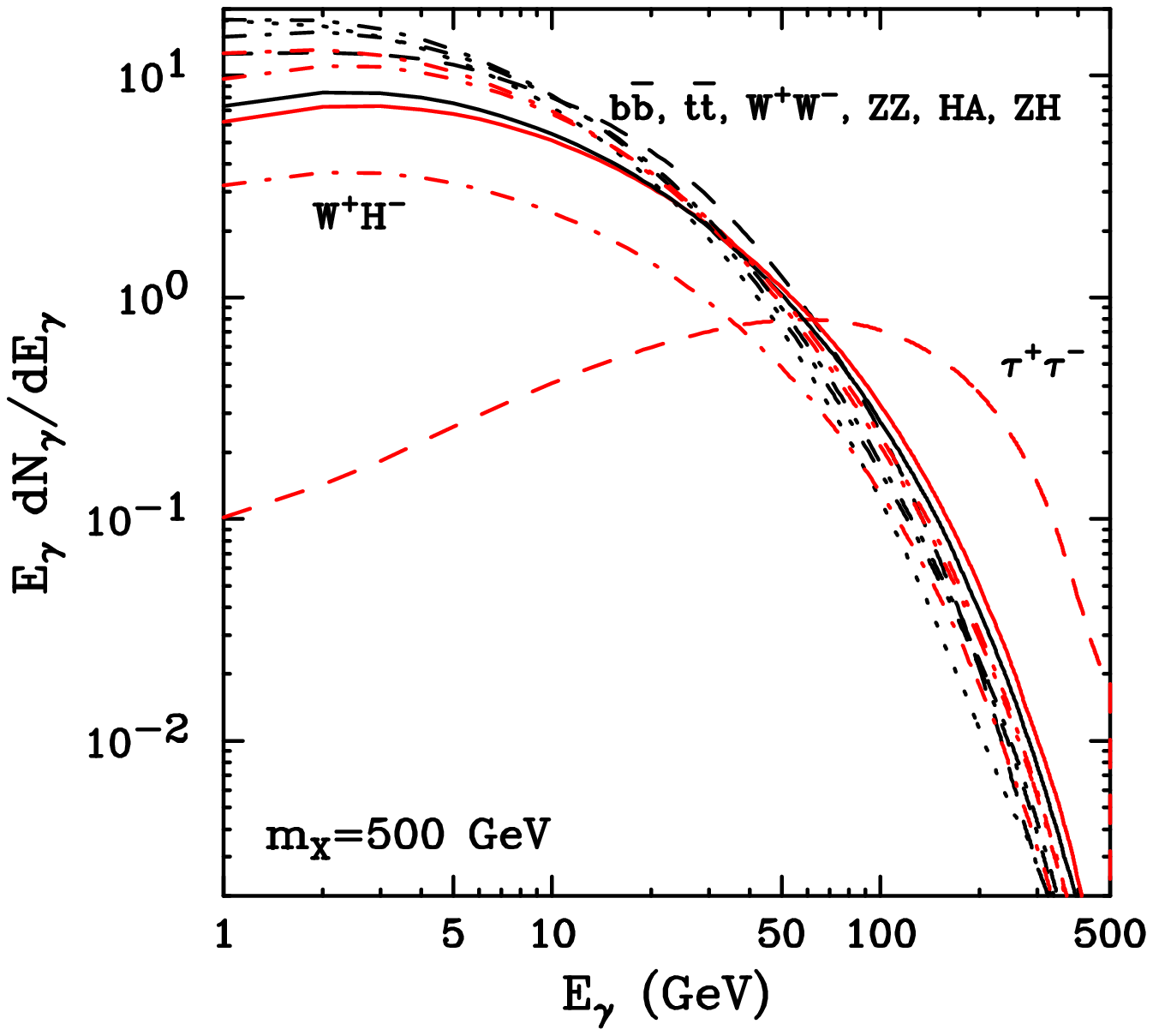}}
\caption{The gamma ray spectrum per annihilation for a 100 GeV (top) and 500 GeV (bottom) WIMP. Each curve denotes the result for a different dominant annihilation mode.}
\label{spectra}
\end{figure}

The energy and angular dependent flux  of gamma rays produced in dark matter annihilations is described by
\begin{equation}
\Phi^{\rm DM} =  \frac{\d N_{\gamma}}{\d E_{\gamma}} \frac{\sigv}{8\pi m^2_X} \int_{\rm{los}} \rho^2(r) \d l,
\label{flux1}
\end{equation}
where $\sigv$ is the WIMP annihilation cross section multiplied by the relative velocity of the two WIMPs 
(averaged over the WIMP velocity distribution), $m_X$ is the mass of the WIMP, $\psi$ is the angle observed relative 
to the direction of the Galactic Center, $\rho(r)$ is the dark matter density as a function of distance to the Galactic 
Center, and the integral is performed over the line-of-sight. $\d N_{\gamma}/\d E_{\gamma}$ is the gamma ray spectrum generated per WIMP annihilation. 
The spectrum of gamma rays produced through dark matter annihilations depends on the nature of the WIMP. Neutralinos, for example, typically annihilate to final states consisting of heavy fermions ($b \bar{b}$, $t \bar{t}$, $\tau^+ \tau^-$) 
or gauge and/or Higgs bosons ($ZZ$, $W^+ W^-$, $HA$, $hA$, $ZH$,
$Zh$, $ZA$, $W^{\pm} H^{\pm}$, where $H$, $h$, $A$ and $H^{\pm}$ are
the Higgs bosons of the Minimal Supersymmetric Standard Model)~\cite{jungman}.
With the exception of the $\tau^+ \tau^-$ channel, each of these annihilation modes result in a very similar spectrum of gamma rays. In Fig.~\ref{spectra}, we show the predicted gamma ray spectrum, per annihilation, for several possible WIMP annihilation modes. In this article, we do not consider mono-energetic gamma ray lines~\cite{lines}, as they are expected to produce far fewer events in GLAST than continuum emission.

In convenient units, Eq.~(\ref{flux1}) can be recast as: 
\begin{eqnarray}
\Phi_{\gamma}(E_{\gamma},\psi) \approx 2.8 \times 10^{-10} \, {\rm cm}^{-2} \, {\rm s}^{-1}\, {\rm sr}^{-1}  \,\,\,\,\,\,\,\,\,\,\,\,\,\,\,\,\,\,\,\,\nonumber \\
\times \frac{dN_{\gamma}}{dE_{\gamma}} \bigg(\frac{\sigv}{3 \times 10^{-26} \,\rm{cm}^3/\rm{s}}\bigg)  \bigg(\frac{100 \, \rm{GeV}}{m_{\rm{X}}}\bigg)^2 
\times J(\psi) \,
\label{flux2}
\end{eqnarray}
where the dimensionless function $J(\psi)$ depends only on the dark matter distribution in the halo
and is defined by convention as:
\begin{equation}
J(\psi) = \frac{1}{8.5 \, \rm{kpc}} \bigg(\frac{1}{0.3 \, \rm{GeV}/\rm{cm}^3}\bigg)^2 \, \int_{\rm{los}} \rho^2(r(l,\psi)) {\rm d} l\,.
\label{jpsi}
\end{equation}
To calculate $J(\psi)$, a model for the dark matter halo distribution must be adopted. A commonly used parameterization of halo profiles is given by 
\begin{equation}
\rho(r) = \frac{\rho_0}{(r/R)^{\gamma} [1 + (r/R)^{\alpha}]^{(\beta - \gamma)/\alpha}} \,,
\label{profile}
\end{equation}
where $R \sim 20$ kpc is the scale radius and $\rho_0$ is fixed by imposing that the dark matter density at the distance of the Sun from the Galactic Center is equal to 0.3 GeV/cm$^3$. Among the most frequently used models is the Navarro-Frenk-White (NFW) profile, which is described by $\alpha = 1$, $\beta=3$ and $\gamma =1 $~\cite{nfw}. When considering the region of the Galactic Center, the most important feature of the halo profile is the inner slope, $\gamma$, which can be steeper or harder
than the assumed fiducial value $\gamma =1$. For example, the Moore {\it et al.} profile is described by $\alpha=1.5$, $\beta=3$, $\gamma=1.5$~\cite{moore}. Note that for  $\gamma\geq 1.5$,  the integral in Eq.~(\ref{jpsi}) diverges. To avoid this behavior in the case of the Moore {\it et al.} profile, we impose a flat core within $10^{-2}$ pc of the Galactic Center.

Although we will use the NFW and Moore {\it et al.} halo profiles as benchmarks, they certainly do not exhaust all possibilities. For a number of reasons, it is quite difficult to predict the dark matter distribution in the inner parsecs of the Galaxy, in which we are most interested. The resolution of N-body simulations is limited to scales of approximately $\sim 10^{2}$ parsecs or so. Furthermore, the gravitational potential in the inner region of the Milky Way is dominated not by dark matter, but by baryons, whose effects are not included in such simulations. The precise impact of baryons on the dark matter distribution is difficult to predict, although an enhancement in the dark matter annihilation rate due to adiabatic compression is generally expected~\cite{ac}. The adiabatic accretion of dark matter onto the central supermassive black hole may also lead to the formation of a density spike in the dark matter distribution. Such a spike would likely result in a very high dark matter annihilation rate~\cite{spike}.

The remaining factors in Eq.~(\ref{flux2}) depend on the particle physics of the dark matter candidate, {\it i.e.} the mass, cross section, and spectrum per annihilation.  As a benchmark value, we adopt a WIMP annihilation cross section of $\sigv \approx 3 \times 10^{-26} \,\rm{cm}^3/\rm{s}$. This is motivated by the fact that a WIMP annihilating with such a cross section during the freeze-out epoch will be generated as a thermal relic with a density similar to the measured dark matter abundance~\cite{wmap}. If these annihilations take place largely through $S$-wave processes, then the annihilation cross section of WIMPs in the Galactic halo ({\it ie.} in the low velocity limit) will also be approximately equal to this value. The annihilation cross section could be considerably reduced, however, if $P$-wave processes or coannihilations between the WIMP and other particles are significant during the freeze-out process.

\section{Projected Constraints}\label{results}

In order to evaluate the ability of GLAST to identify or constrain the properties of annihilating dark matter, we consider the spatial and spectral information simultaneously. We use the model introduced in the previous sections to construct a simulated sky map, free of a dark matter signal. Then, for given dark matter mass and annihilation cross section, we compute the $\chi^2$ of the simulated sky against models including a contribution from dark matter annihilation radiation. We then repeat the analysis against the same simulated sky for a wide range of models.

The simulated sky contains the two resolved sources and the diffuse background described in Section~\ref{backgrounds}. For a given angular pixel, $\nu$, and energy bin, $k$, we calculate the total number of expected photons:
\begin{equation}
O_{\nu, k}^{\rm obs}=B_{\nu, k}^{\rm ACT}+B_{\nu, k}^{\rm EG}+B_{\nu, k}^{\rm diff}(A,\alpha),\,\label{allback}
\end{equation}
and then add in Poisson noise. For a given running time of the experiment, $t$, each of these terms is obtained by integrating the flux over an given energy and angular bin,
\begin{equation}
B_{\nu, k}^{(i)}=t\times \epsilon\times \int_{\Delta \Omega_\nu}{\rm d}\Omega \int_{\Delta E_k} {\rm d}E\, A_{\rm eff}\,\Phi^{(i)}\,\frac{\d{\mathcal P}}{\d \Omega}.
\end{equation}
Each angular bin covers solid angle $\Delta\Omega_\nu=0.01^{\circ} \times 0.01^{\circ}$ (smaller than the angular resolution of GLAST). Since we consider the inner $2^{\circ} \times 2^{\circ}$ field-of-view, there are a total of 40,000 bins. We include 25 energy bins distributed logarithmically between 1 and 300 GeV (approximately 10 bins per energy decade, which is realistic given GLAST energy resolution).
 
The distribution over the sky, $O_{\nu, k}^{\rm obs}$, therefore, does not contain any contribution from dark matter annihilations. We compare this to a model which includes both the backgrounds and a signal from dark matter:
\begin{equation}
R_{\nu, k}=\,S_{\nu, k}(m_X,\sigv)+B_{\nu, k}^{\rm ACT}+B_{\nu, k}^{\rm EG}+B_{\nu, k}^{\rm diff}(A',\alpha')\,.
\end{equation}
We do this by computing the $\chi^2$ of the fit as follows~\footnote{Strictly speaking, this expression is valid only if several events are present in a given energy/angular bin. This is not always the case in the present application. The most important bins, however, which contribute most to the discrimating power of this calculation, do contain a sufficient number of events. We have checked this by increasing the size of our angular bins and find good agreement.}:
\begin{equation}
\chi^2\left(A^{\rm ACT}, A^{\rm EG},\alpha',m_X,\sigv\right) = \sum_{\nu,k}
\frac{ \left( R_{\nu,k}-O_{\nu,k}^{\rm obs}\right)^2}  {O_{\nu,k}^{\rm obs}}.
\end{equation}
The free parameters in the $\chi^2$ are the slope of the diffuse background, the amplitudes of the point source backgrounds, and the mass and annihilation cross section of the dark matter particle. We fix the amplitude of the diffuse spectrum by summing over all events. A more complete analysis would marginalize over this amplitude as well, but  we are in the limit where the overall number of photons is dominated by the astrophysical background.
Fixing the quantity reduces the CPU time required while leading to only a small underestimate of the errors on the dark matter parameters.

Although we have included only the HESS and EGRET sources (in addition to the diffuse background) in our analysis, this method could easily be extended to include any other astrophysical point sources to be discovered by GLAST. Also, in principle one could have introduced additional parameters in the modeling of the point-like sources. Computational constraints, however, require us to limit the number of free parameters in our analysis, and the most difficult background to separate is the diffuse component. Concerning the point-like source at the Galactic Center, it is worth mentioning that the bulk of the statistical significance of the dark matter annihilation signal does not come from the inner $0.1^{\circ}$ around the Galactic Center (where the HESS source dominates), but rather from the surrounding angular region, even in the case of a cusped halo profile~\cite{glasthaze}.

\begin{figure}
\resizebox{8.5cm}{!}{\includegraphics{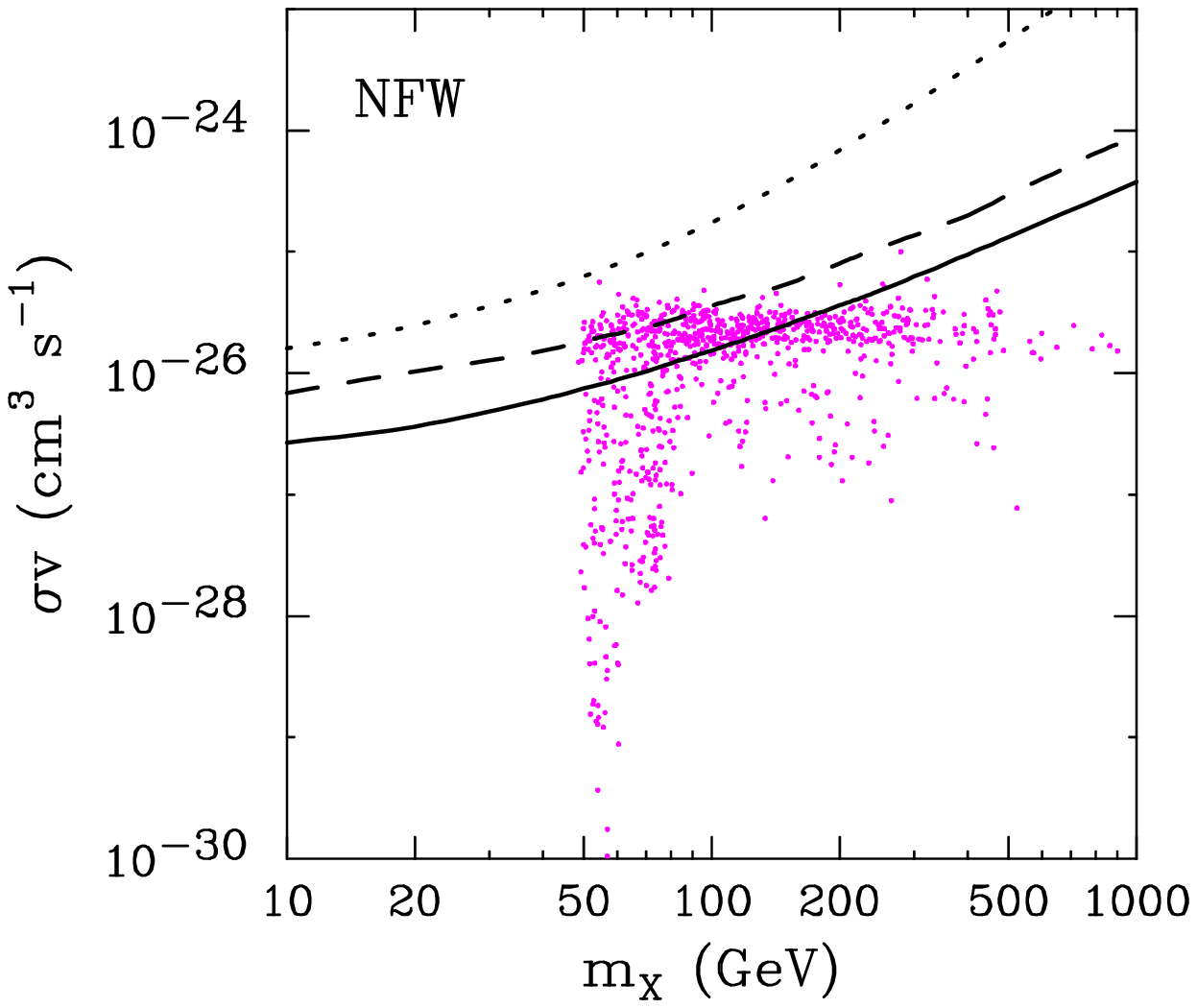}}
\resizebox{8.5cm}{!}{\includegraphics{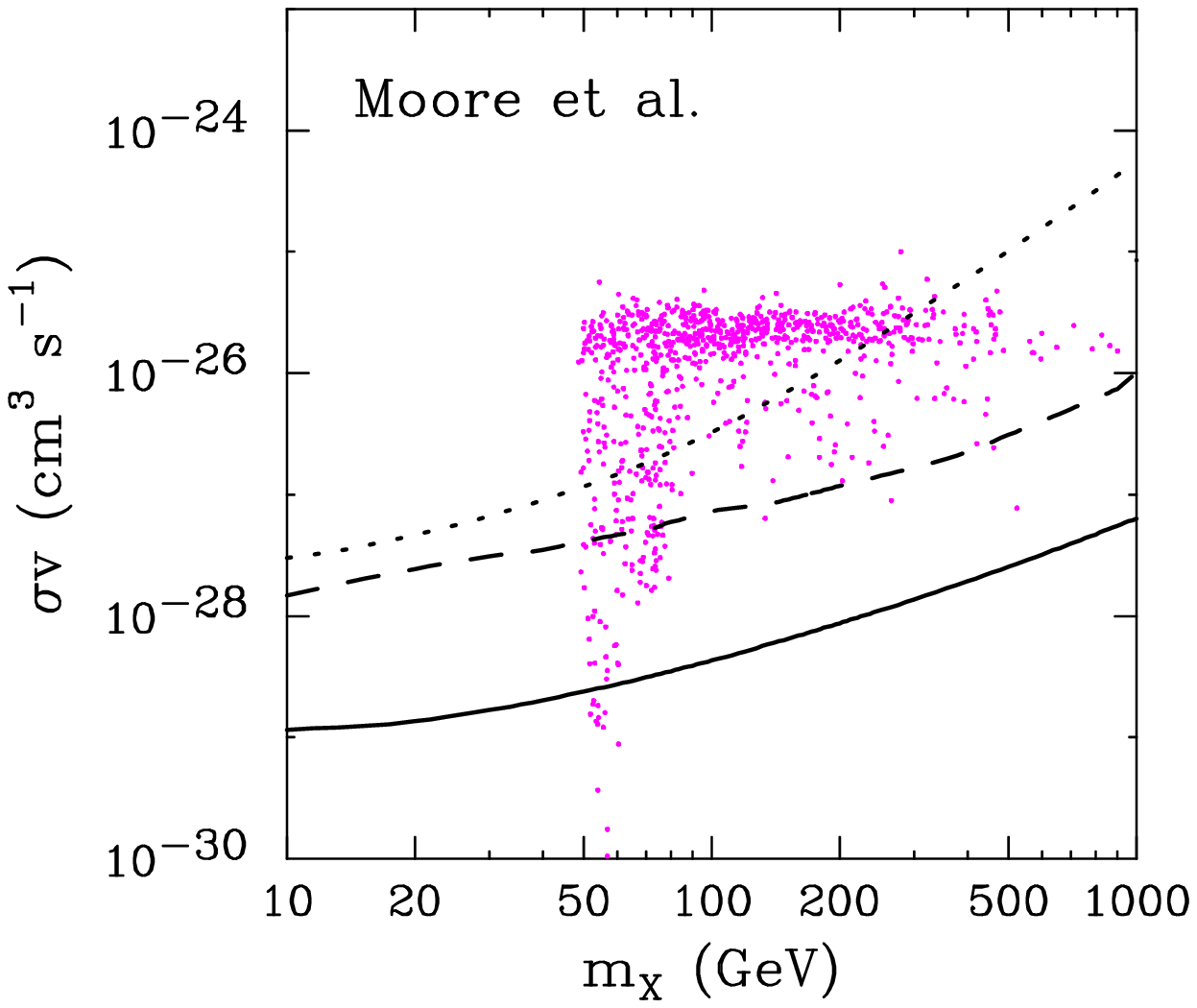}}
\caption{The projected exclusion limits at 95\% confidence level from GLAST (after ten years) on the WIMP annihilation cross section,  as a function of the WIMP mass. The region above the dotted line is already excluded by EGRET~\cite{dingus}. The dashed and solid lines show the projections
for GLAST for an assumed isotropic diffuse background and the limit case where the astrophysical background has exactly the same angular distribution of the DM signal, respectively.  In the upper and lower frames, the NFW and Moore {\it et al}.~halo profiles have been adopted, respectively.  Also shown are points representing a random scan of supersymmetric models.}
\label{limitnfw}
\end{figure}

By scanning over this five-dimensional parameter space, we can evaluate how large a contribution from dark matter annihilations can be 
contained in $R_{\nu, k}$ before the total signal becomes inconsistent with the background-only sky, $O_{\nu, k}^{\rm obs}$. 
In Fig.~\ref{limitnfw}, we show the projected exclusion limits at the 95\% confidence level in the $\{m_X,\,\sigv\}$  plane for ten years of collection time by GLAST 
and for the cases of an NFW (top frame) or Moore {\it et al.} (bottom frame) halo profile and WIMP annihilations to $b \bar{b}$. 
The solid line in each frame represents the limit found if the diffuse background is assumed to be distributed isotropically, while the dashed line represents the conservative limit obtained if the diffuse background has the same angular distribution as the dark matter signal ({\it i.e.} the case in which angular information is not useful in disentangling the signal from the diffuse background). For values of $\sigv$ below the corresponding lines, a pure background model of the kind of Eq.~(\ref{allback}) is expected to be consistent with the data.
The fact that the limits are significantly stronger in the uniform background case is the manifestation of the
improved sensitivity which can be achieved by an analysis including both energy and angular information.
 For comparison,  in Fig.~\ref{limitnfw}  we also show the region already excluded by EGRET~\cite{dingus} (above the dotted line) and the mass and cross section of neutralino models found in a random scan over supersymmetric parameters, as calculated using DarkSUSY~\cite{darksusy}. As expected, many of the models cluster around $\sigv \sim 3 \times 10^{-26}$ cm$^3$/s, the value required of a thermal relic annihilating via an $S$-wave amplitude. Each point shown represents a model which respects all direct collider constraint and generates a thermal dark matter abundance consistent with the observed dark matter density. In our scan, we have varied the SUSY parameters $M_2$, $|\mu|$ and $m_{\tilde{q}}$ up to 2 TeV, $m_A$ and $m_{\tilde{l}}$ up to 1 TeV and $\tan \beta$ up to 60. We have assumed the gaugino masses evolve to a single unified scale, such that $M_1 \approx 0.5 M_2$, $M_3 \approx 2.7 M_2$.

\begin{figure}
\resizebox{8.5cm}{!}{\includegraphics{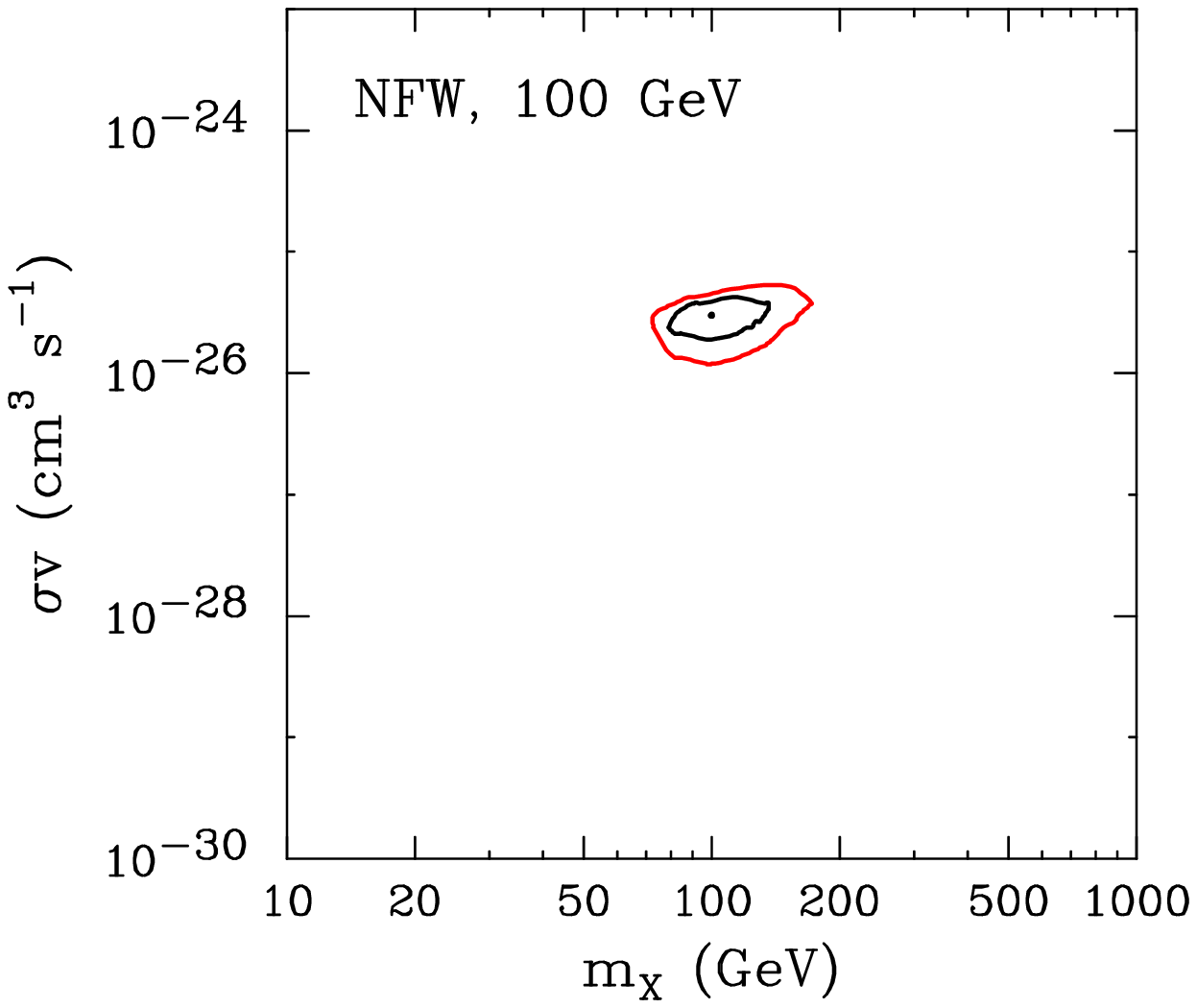}}
\resizebox{8.5cm}{!}{\includegraphics{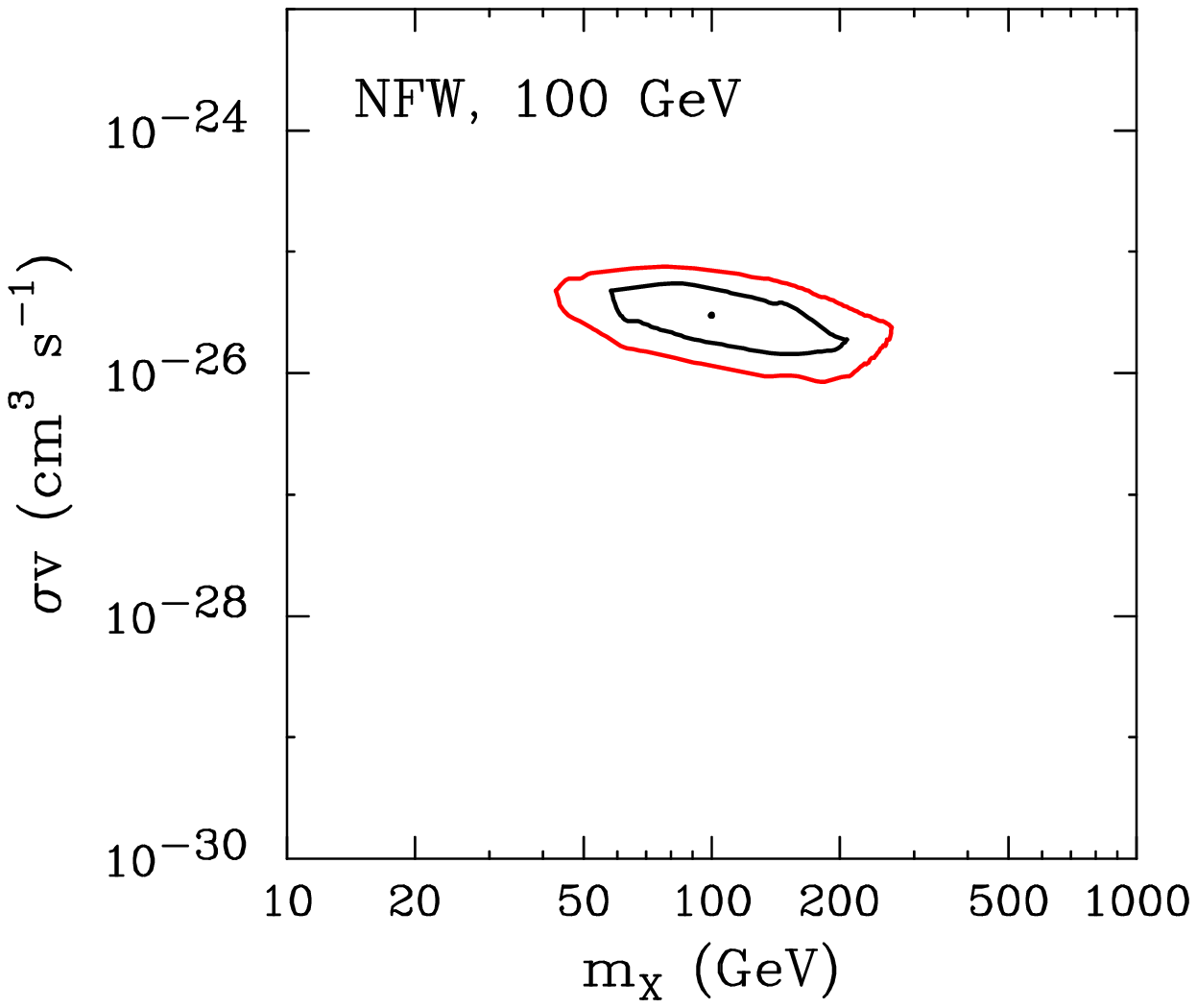}}
\caption{The ability of GLAST to measure the annihilation cross section and mass of dark matter after ten years of observation. Here, we have used a benchmark scenario with $m_X=100$ GeV, $\sigv = 3 \times 10^{-26}$ cm$^3$/s and an NFW halo profile. The inner and outer contours in each frame represent the 2 and 3$\,\sigma$ regions, respectively. In the top frame, the halo profile shape was treated as if it is known in advance. In the lower frame, we marginalize over the inner slope of the profile. }
\label{ellipsenfw}
\end{figure}

\section{Measuring The Properties of Dark Matter}\label{measure}

Once gamma rays are identified as having been produced in dark matter annihilations, such observations could then be used to measure the characteristics of the dark matter particle, including its mass, annihilation cross section and spatial distribution. Such determinations are an important step toward identifying the particle nature of dark matter. In this section, we discuss GLAST's ability to constrain these properties.

To accomplish this, we do a similar calculation to that performed in Sec.~\ref{results}, but now also include a contribution from dark matter annihilations in the quantity, $O^{\rm obs}_{\nu, k}$. In particular, we include the contribution from WIMPs distributed with an NFW halo profile, with an annihilation cross section of $3 \times 10^{-26} \,\rm{cm}^3/\rm{s}$ and a mass of 100 GeV. We then calculate the statistical significance at which these properties can be measured by GLAST.

\begin{figure}
\resizebox{8.5cm}{!}{\includegraphics{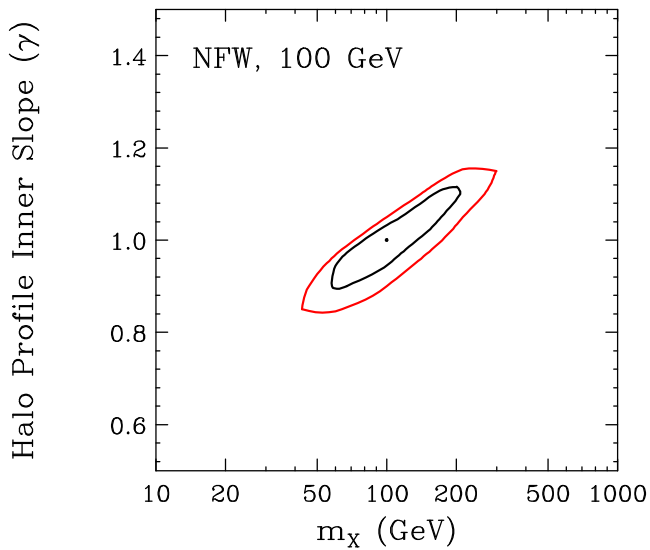}}
\caption{The ability of GLAST to measure the inner slope of the halo profile and the mass of dark matter particle (marginalizing over the annihilation cross section) after ten years of observation. Here, we have used a benchmark scenario with $m_X=100$ GeV, $\sigv = 3 \times 10^{-26}$ cm$^3$/s and an NFW halo profile. The inner and outer contours represent the 2 and 3$\,\sigma$ regions, respectively.}
\label{ellipseslope}
\end{figure}

In Fig.~\ref{ellipsenfw}, we show the ability of GLAST to determine the WIMP mass and annihilation cross section in this case. In each frame, we show the projected 2 and 3$\,\sigma$ constraints on these properties as may be determined by GLAST,  assuming an isotropic diffuse background (in addition to background point sources). In the top frame, we treat the shape of the halo profile (NFW) as if it is known in advance. Of course, this is not a realistic assumption, and a less accurate determination of the WIMP mass must be expected in a more realistic treatment. In the lower panel of Fig.~\ref{ellipsenfw} we report the results obtained marginalizing over the inner slope of the halo profile, $\gamma$ [see Eq.~(\ref{profile})]. Without a prior on the inner halo slope,
the constraint on the dark matter mass worsens by a factor $\sim $2. 

If the spectrum and angular distribution of gamma rays from dark matter annihilations in the Galactic Center region are sufficiently well measured, it will also be possible to measure the underlying dark matter distribution. In Fig.~\ref{ellipseslope}, we project the results from the lower frame of Fig.~\ref{ellipsenfw} in the  $\{m_X,\,\gamma\}$ plane (marginalized over the annihilation cross section).  In our benchmark model, the inner slope of the halo profile can be determined at approximately the $\sim 10\%$ level.

\section{Discussion and Conclusions}\label{conclusions}

The most challenging task for indirect dark matter searches is not detecting the products of dark matter annihilations, but in confidently identifying those particles as such. In particular, any signal must be separated from astrophysical backgrounds if it is to be reliably claimed to be a detection of dark matter annihilation products. This is certainly true in the case of gamma ray telescopes hoping to observe dark matter annihilations in the region of the Galactic Center, where astrophysical backgrounds are especially foreboding.

In this paper, we have studied quantitatively the ability of GLAST to identify gamma rays from dark matter annihilations in the Galactic Center region, using all information encoded in the spectrum and angular distribution of the observed events. Due to the complex nature of the Galactic Center's backgrounds and dark matter distribution, it is important for GLAST to make full use of this information to best separate dark matter annihilation products from any astrophysical backgrounds which are present.

When GLAST begins its mission in 2008, it will detect a number of astrophysical sources in the region of the sky around the Galactic Center. These include the point sources identified by HESS and EGRET, and perhaps other, thus far unknown, sources. A diffuse gamma ray background will also likely be present. In this article, we have modeled these backgrounds based on known spectral information, and using a point-spread-function for GLAST, in the hope of identifying statistically significant departures from this model resulting from dark matter annihilations.

The spectrum of gamma rays from dark matter annihilations is quite distinctive, and does not resemble the power-law form observed from typically astrophysical sources. Furthermore, the angular distribution of dark matter annihilation products is not a simple point source, nor is it isotropic or trace the Galactic Ridge, as the backgrounds are expected to. By exploiting these differences, it may be possible to  identify the products of dark matter annihilation observed from the inner galaxy, even they are a subdominant component of the total emission.

If dark matter annihilation products are identified by GLAST, then it may also be possible to measure or constrain the properties of dark matter, including its mass, annihilation cross section and spatial distribution.  We find that GLAST is unlikely to determine the WIMP's mass with high precision, however. For example, for the case of a 100 GeV WIMP with an annihilation cross section of $3 \times 10^{-26}$ cm$^3$/s and distributed with an NFW halo profile, the mass could be determined to lie within approximately 50-300 GeV. In the same benchmark model, the inner slope of the dark matter halo profile could be determined to $\sim 10\%$ precision by GLAST.

{\it Note added:} After we had completed this study, we became aware of updated estimates for the performance of GLAST~\cite{lat}. In particular, the point spread function that we have used is a factor $3 \div 4$ more optimistic than was found in the current estimate. As a consequence, a reduction of the ability of GLAST to identify dark matter and to measure its properties is expected, but not to an overwhelming extent: we checked for example that for a NFW profile and a DM mass of 100 GeV the bound worsens by less than 30\%. Also, additional effects not taken into account here (like effective
area degradation due to orbit inclination
and spacecraft rocking; non-negligible dead-time and a high-particle-background associated to orbital passage over the South-Atlantic Anomaly) may reduce the exposure by up to a factor $\sim 2$ compared to the present estimate. We would like to thank the referee for bringing this to our attention.

\bigskip
{\bf Acknowledgments} This work has been supported by the US Department of Energy and by NASA grant NAG5-10842. Fermilab is operated by Fermi Research Alliance, LLC under Contract No.~DE-AC02-07CH11359 with the United States Department of Energy.

\end{document}